\documentstyle[psfig]{l-aa}

\newcommand{\aaa}[2]{A\&A #1, #2}

\newcommand{\aj}[2]{AJ #1, #2}
\newcommand{\apj}[2]{ApJ #1, #2}
\newcommand{\apjs}[2]{ApJS #1, #2}

\newcommand{\mnras}[2]{MNRAS #1, #2}
\newcommand{\nat}[2]{Nat #1, #2}

\newcommand{\rmp}[2]{Rev. Mod. Phys. #1, #2}
\newcommand{\sval}[2]{SvAL #1, #2}

\begin{document}

\thesaurus{04 (02.07.1;
               03.13.1;
               03.13.4;
               11.05.2;
               11.11.1;
               11.19.2)}

\title{Modelling gravity in $N$-body simulations of disc galaxies}

\subtitle{Optimal types of softening for given dynamical requirements}

\author{Alessandro B. Romeo}

\institute{Onsala Space Observatory,
           Chalmers University of Technology,
           S-43992 Onsala, Sweden
           (romeo@oso.chalmers.se)}

\date{Received 6 March 1998 / Accepted 21 April 1998}

\maketitle

\begin{abstract}

Modelling gravity is a fundamental problem that must be tackled in $N$-body
simulations of stellar systems, and satisfactory solutions require a deep
understanding of the dynamical effects of softening. In a previous paper
(Romeo 1997), we have devised a method for exploring such effects, and we
have focused on two applications that reveal the dynamical differences
between the most representative types of softened gravity. In the present
paper we show that our method can be applied in another, more fruitful, way:
for developing new ideas about softening. Indeed, it opens a {\it direct\/}
route to the discovery of optimal types of softened gravity for given
dynamical requirements, and thus to the accomplishment of a physically
consistent modelling of disc galaxies, even in the presence of a cold
interstellar gaseous component and in situations that demand anisotropic
resolution.


\end{abstract}

\section{Introduction}

$N$-body simulations of disc galaxies rely on the use of softening. This
artifice removes the short-range singularity of the gravitational
interaction, which is dynamically unimportant and computationally
troublesome, whereas it leaves the long-range behaviour of gravity unchanged.
But softening is also a critical factor in simulations. It controls their
quality and can affect their result on scales much larger than the softening
length. Its dynamical effects are further exacerbated in the presence of a
cold interstellar gaseous component and in situations that demand anisotropic
resolution. Thus softening poses a dynamical problem of special concern,
which should be probed carefully and in detail (e.g., Hernquist \& Barnes
1990; Pfenniger \& Friedli 1993; Romeo 1994, hereafter Paper I; Romeo 1997,
hereafter Paper II%
\footnote{Sections and equations of that paper are denoted by the prefix
          II.};
and references therein).

   In Paper I, we have investigated how faithful simulations are. In
particular, we have concluded that the standard way of introducing softening
in the presence of stars and cold interstellar gas is definitely
unsatisfactory in several regimes of astrophysical interest. It is so because
important small-scale instabilities of the gaseous component, e.g.\ those
peculiar to star-formation processes, are suppressed just as unphysical noise
of the stellar component. Faithfulness requires an appropriate introduction
of two softening lengths, one for each component, and also a rigorous
specification of the star-gas gravitational interaction.

   In Paper II, we have devised a method for exploring the dynamical effects
of softening. As a major result, we have shown how to choose the softening
length for optimizing the faithfulness of simulations to the Newtonian
dynamics. Then we have focused on two applications that reveal the dynamical
differences between the most representative types of softened gravity. In
particular, we have concluded that it is desirable to improve the current way
of introducing anisotropic softening. We need a clearer decoupling of the
resolution parallel and perpendicular to the plane, and also more natural
planar and vertical softening lengths.

   In the present paper, which completes our planned research work about
softening, we propose an {\it innovative\/} solution to the problem. The
understanding of galactic and extragalactic astrophysics is at a crucial
stage. Unsolved problems are viewed in new perspectives, which promise major
revisions of knowledge (see, e.g., Blitz \& Teuben 1996; Block \& Greenberg
1996). Recent investigations suggest, for instance, a more enigmatic
interplay between stellar disc and bulge/halo (e.g., Lequeux et al.\ 1995), a
clearer relation between cold gas and dark matter in spiral galaxies (e.g.,
Pfenniger et al.\ 1994; Pfenniger \& Combes 1994; Combes \& Pfenniger 1997),
and a closer connection between the fractal structures of the interstellar
medium and of the universe (e.g., de Vega et al.\ 1996, 1998). The
implications are clear: modelling gravity in $N$-body simulations of disc
galaxies should offer a flexible interface with such a progress. Our solution
is to optimize the fidelity of simulations to given dynamical requirements.
How do we apply this idea in practice?
\begin{enumerate}
\item We impose the requirements in the wavenumber space since this is the
      natural dynamical domain of gravity, as Pfenniger \& Friedli (1993)
      have previously emphasized.
\item We identify the softening length with the characteristic dynamical
      scale length.
\item Then we invert part of the method of Paper II, and the result is the
      optimal type of softened gravity that satisfies those dynamical
      requirements.
\end{enumerate}
Our application covers both 2-D and 3-D modelling. The basic cases are
extended to more complex situations through recipes for implementing star-gas
and anisotropic softening, which have already been motivated (cf.\
discussions of Papers I and II). Last but not least, each description is
complemented by an example that leaves room for creativity.

   The present paper is organized as follows. The application is shown in
Sects.\ 2 and 3 (see also Appendix A), and proceeds as in the previous
discussion. Comments on related works concerning softening are made in Sect.\
4. The conclusions and perspectives are drawn in Sect.\ 5, where we present
our three papers about softening in a more unified view and emphasize their
potentially strong impact on galactic dynamics.


\begin{figure*}
\vbox{\vspace{.1cm}
      \hbox{\hspace{-.25cm}
            \psfig{figure=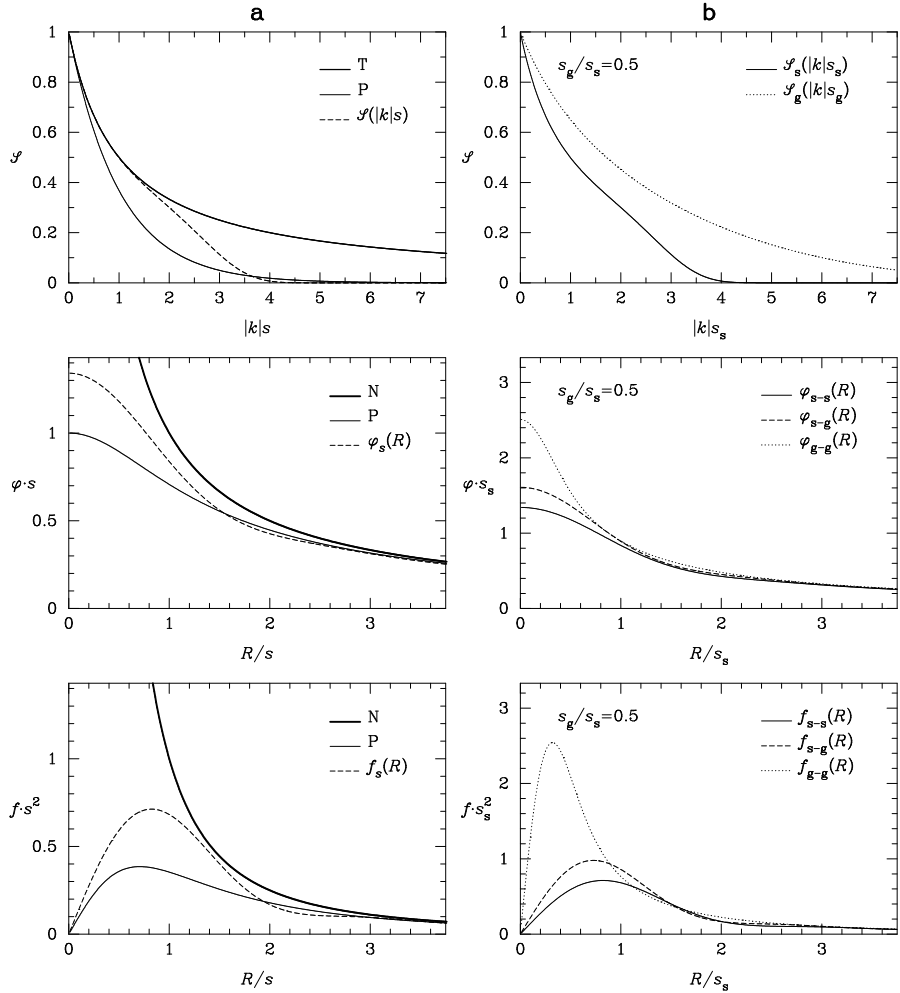,width=18.9cm,height=13.275cm,angle=-90}}
      \vspace{-2.9cm}}
\hfill\parbox[b]{5.7cm}{\caption[]{Examples of 2-D modelling: {\bf a}
                                   one-com\-pon\-ent case (cf.\ Sect.\ 2.1),
                                   {\bf b} two-com\-pon\-ent case (cf.\
                                   Sect.\ 2.2). The abbreviations N, T and P
                                   mean Newtonian gravity, thickness and
                                   Plummer softening, respectively}}
\end{figure*}


\begin{figure*}
\vbox{\vspace{.1cm}
      \hbox{\hspace{-.25cm}
            \psfig{figure=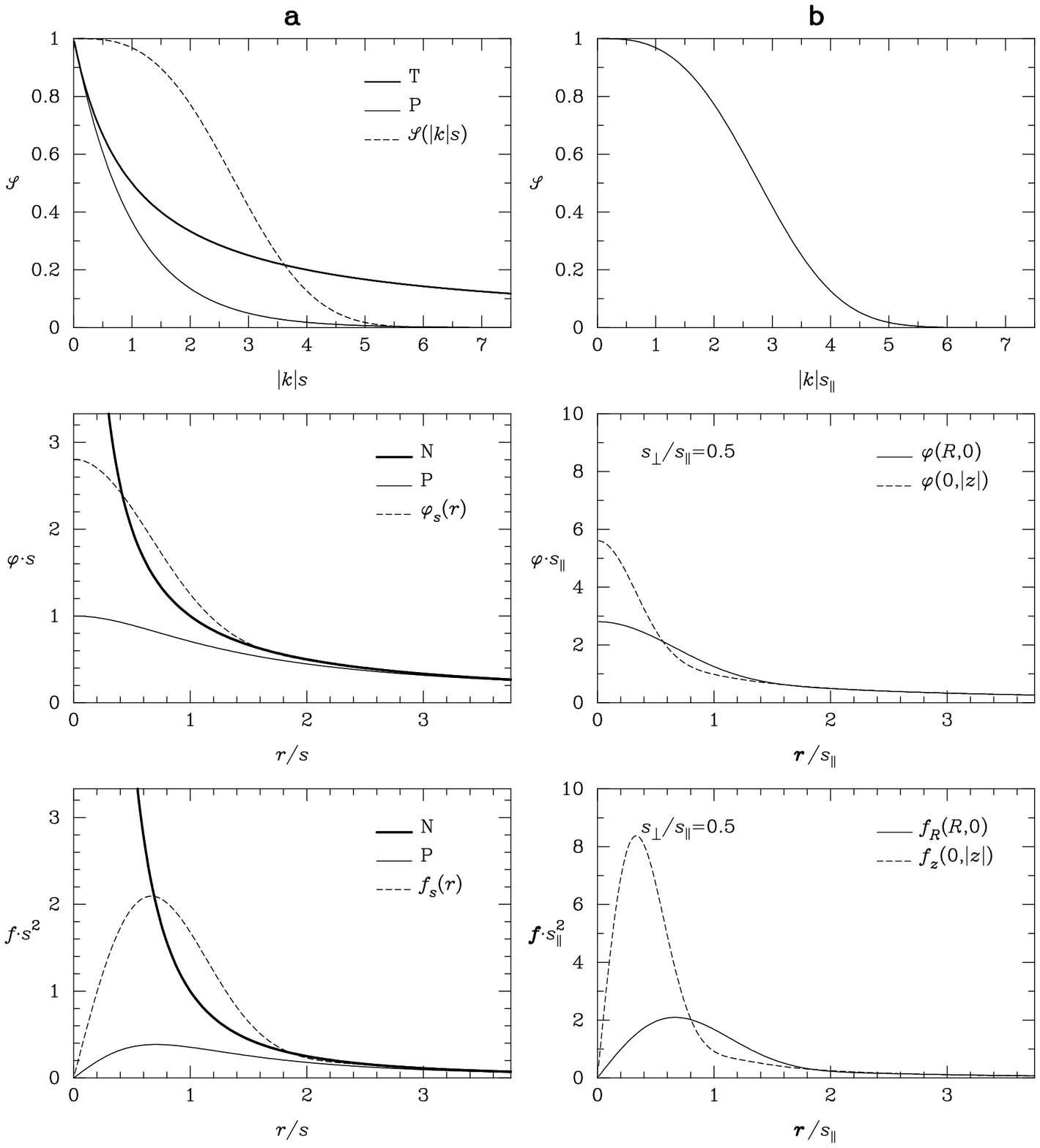,width=18.9cm,height=13.275cm,angle=-90}}
      \vspace{-2.9cm}}
\hfill\parbox[b]{5.7cm}{\caption[]{Examples of 3-D modelling: {\bf a}
                                   iso\-tropic case (cf.\ Sect.\ 3.1), {\bf
                                   b} an\-iso\-tropic case (cf.\ Sect.\ 3.2).
                                   The abbreviations N, T and P mean
                                   Newtonian gravity, thickness and Plummer
                                   softening, respectively}}
\end{figure*}


\section{2-D modelling}

\subsection{Inverting part of the method of Paper II}

The method of Paper II allows determining the dynamical response of the model
to a given type of softened gravity. The basic quantity that describes this
response is the reduction factor ${\cal S}(|k|s)$ defined in Eq.\ (II-4),
where $k$ has the meaning of a radial wavenumber and $s$ is the softening
length. In a few words, this is the factor by which softening reduces the
dynamical contribution of self-gravity. The behaviour of ${\cal S}(|k|s)$
provides the following information:
\begin{itemize}
\item At small $|k|s$, it shows how significantly the stability and
      collective relaxation properties are affected on large scales.
      Specifically, a comparison with the reduction factor of 3-D discs with
      Newtonian gravity reveals how well softening mimics the effects of
      thickness, as far as density waves are concerned.
\item At large $|k|s$, it shows how effectively noise is suppressed on small
      scales.
\item It also contains less direct information about how significantly the
      equilibrium properties are affected.
\end{itemize}
The reduction factor ${\cal S}(|k|s)$ is related to the point-mass potential
$-Gm\varphi_s(R)$ and force $-Gm^2f_s(R)$ through integral transforms. These
transforms can easily be inverted, and the inversion formulae that relate
$\varphi_s(R)$ and $f_s(R)$ to ${\cal S}(|k|s)$ are:
\begin{equation}
\varphi_s(R)={\cal H}_0\left[\frac{1}{k}\,{\cal S}(|k|s)\right](R)\, ,
\end{equation}
\begin{equation}
f_s(R)={\cal H}_1[{\cal S}(|k|s)](R)\, ,
\end{equation}
where ${\cal H}_\nu$ denotes the Hankel transform of order $\nu$:
\begin{equation}
{\cal H}_\nu[g(k)](R)=\int_0^\infty\!g(k)\,{\rm J}_\nu(kR)\,k\,{\rm d}k\, ,
\end{equation}
and ${\rm J}_\nu$ denotes the Bessel function of the first kind and order
$\nu$ (for mathematical and numerical references see Sect.\ II-2.1.1; in
particular, for a swift numerical computation of Hankel transforms use, e.g.,
the NAG library). Inverting this part of the method has a direct practical
importance and, indeed, is the trick behind the present application: it
allows imposing {\it arbitrary\/} dynamical requirements in the form of a
reduction factor and identifying the optimal type of softened gravity that
satisfies such requirements. In addition, the rest of the method allows
testing the precise dynamical performance of the modelling.

   Here is an example: we want to find a type of softened gravity that mimics
the effects of thickness very well and, at the same time, suppresses noise
very effectively. This is illustrated in Fig.\ 1a and in the following
discussion. A reduction factor that conforms with the dynamical requirements
specified above is:
\begin{equation}
{\cal S}(|k|s)=\frac{1}{1+|k|s}\cdot{\rm e}^{-(|k|s/\pi)^5}\, .
\end{equation}
The first function is the reduction factor of 3-D discs with Newtonian
gravity and characteristic scale height $s$ (Shu 1968; Vandervoort 1970;
Romeo 1992). The second function acts as a low-pass filter with rather sharp
cut-off at radial wavelengths $\lambda\approx 2s$. The benefits of filtering
in spectral domain are well known in the context of digital image processing
(see, e.g., Jain 1989; Press et al.\ 1992; see also Vetterling et al.\ 1992).
In our case the dynamical resolution, i.e.\ the faithfulness in simulating
the Newtonian dynamics, is substantially higher than in the standard Plummer
softening. Indeed, it can be further improved by choosing sharper filters, at
the cost of noticeable oscillations in the gravitational interaction. To some
extent, their preference is a matter of taste. On the other hand, too sharp
filters make the typical radial wavelength and possibly other dynamical
properties hypersensitive to the location of the cut-off, which is
unphysical. Thus, they should not be used.

\subsection{Implementing star-gas softening}

How do we model gravity in the presence of two components, such as stars and
cold interstellar gas? Let us think in the alternative way of finite-sized
particles interacting with Newtonian gravity, as Dyer \& Ip (1993) have
partly suggested (for an ABC of finite-sized particles see Appendix A). Then
the answer is simple. Each component turns out to have its own positive
reduction factor ${\cal S}(|k|s)$, where now $s$ is the scale length of the
particle mass distribution. So the star-star and gas-gas interactions are as
in the one-component case, while the star-gas interaction potential $-Gm_{\rm
s}m_{\rm g}\varphi_{\mbox{\scriptsize s-g}}(R)$ and force $-Gm_{\rm s}m_{\rm
g}f_{\mbox{\scriptsize s-g}}(R)$ are determined unequivocally by the
inversion formulae:
\begin{equation}
\varphi_{\mbox{\scriptsize s-g}}(R)
={\cal H}_0\left[\frac{1}{k}\,
                 \sqrt{{\cal S}_{\rm s}(|k|s_{\rm s})\cdot
                       {\cal S}_{\rm g}(|k|s_{\rm g})}\,\right](R)\, ,
\end{equation}
\begin{equation}
f_{\mbox{\scriptsize s-g}}(R)
={\cal H}_1\left[\sqrt{{\cal S}_{\rm s}(|k|s_{\rm s})\cdot
                       {\cal S}_{\rm g}(|k|s_{\rm g})}\,\right](R)\, .
\end{equation}
Our recipe for implementing star-gas softening has strong advantages. Indeed,
its characteristics are {\it fundamental\/} for modelling the complex roles
that such components play in regimes of astrophysical interest, as we have
concluded in Paper I.

   As an example, we want to generalize the type of softened gravity found in
the one-component case to the presence of a young disc stellar population and
a cold interstellar gaseous component with, say, $s_{\rm s}:s_{\rm g}=2:1$
(see, e.g., Mihalas \& Binney 1981). This is illustrated in Fig.\ 1b and in
the following discussion. The finite-sized particle implementation of
star-gas softening is consistent with the effects of thickness: there are two
positive reduction factors, one for each component. Again, the stellar
reduction factor is:
\begin{equation}
{\cal S}_{\rm s}(|k|s_{\rm s})
=\frac{1}{1+|k|s_{\rm s}}\cdot{\rm e}^{-(|k|s_{\rm s}/\pi)^5}\, .
\end{equation}
Concerning cold interstellar gas, the situation is more complex and uncertain
(e.g., Combes \& Pfenniger 1996; Elmegreen 1996a, b; Ferrara 1996; Lequeux \&
Gu\'elin 1996; Pfenniger 1996; Pfenniger et al.\ 1996). Sharp filters should
not be used because they over-stress adherence to the effects of thickness,
whereas the effects of turbulence and fractality may be more important. Soft
filters are safer in that respect, and our preference goes to the Gaussian
member of the family%
\footnote{A simpler, but less instructive, choice would be: ${\cal S}_{\rm
          g}(|k|s_{\rm g})={\rm e}^{-|k|s_{\rm g}}$, i.e.\ the reduction
          factor for the standard Plummer softening.}.
So the gaseous reduction factor is:
\begin{equation}
{\cal S}_{\rm g}(|k|s_{\rm g})
=\frac{1}{1+|k|s_{\rm g}}\cdot{\rm e}^{-(|k|s_{\rm g}/\pi)^2}\, .
\end{equation}
Regarding $s_{\rm s}$ as the characteristic scale height of reference,
different values of $s_{\rm g}/s_{\rm s}$ have no influence on the stellar
functions, modify the star-gas gravitational interaction moderately and
change the gaseous functions according to simple scaling laws.

\section{3-D modelling}

\subsection{Passing from 2D to 3D}

In 3-D disc models, ${\cal S}(|k|s)$ is no longer the true reduction factor
but is still useful for quantifying the dynamical effects of softening
parallel to the plane, which now combine with those of vertical random
motion. Specifically, a comparison with the reduction factor of 3-D discs
with Newtonian gravity suggests how much softening interferes with the
effects of thickness, as far as density waves are concerned. The inversion
formulae for $\varphi_s(r)$ and $f_s(r)$ are as in the 2-D case. This is the
simplest way of passing from 2-D to 3-D modelling. (The generalization to two
components is clear.)

   Here is an example: we want to find a type of softened gravity that
interferes with the effects of thickness very little and, at the same time,
suppresses noise very effectively. This is illustrated in Fig.\ 2a. The
pseudo reduction factor is:
\begin{equation}
{\cal S}(|k|s)={\rm e}^{-(|k|s/\pi)^3}\, ,
\end{equation}
and acts as a low-pass filter with rather soft cut-off at $\lambda\sim 2s$.
The discussion concerning the dynamical resolution and the action of sharper
filters follows the 2-D case closely. On the other hand, only in 3D can we
simulate the evolutionary nature of thickness, which arises from the vertical
random motion and its subtle coupling with the dynamical properties parallel
to the plane (Romeo 1990, 1992; see also Paper I).

\subsection{Implementing anisotropic softening}

In order to model gravity in situations that demand anisotropic resolution,
it is convenient to think in the standard way of point particles interacting
with softened gravity, as Zotov \& Morozov (1987) have partly suggested. The
softening surface is transformed from a sphere of radius $s$ into a spheroid
of planar and vertical semi-axes $s_\parallel$ and $s_\perp$, respectively.
This means that the softening length is the distance from the centre to the
surface of the spheroid in the direction of the position vector:
\begin{equation}
s(R,|z|)=\sqrt{\frac{s_\parallel^2R^2+s_\perp^2z^2}{R^2+z^2}}\, .
\end{equation}
The resolution turns out to be decoupled parallel and perpendicular to the
plane, and to be determined by the natural planar and vertical softening
lengths. These characteristics of our recipe for implementing anisotropic
softening have important advantages, as we have concluded in Paper II. (The
generalization to two components is clear.)

   As an example, we want to generalize the type of softened gravity found in
the isotropic case to situations that demand moderate anisotropic resolution
with, say, $s_\parallel:s_\perp=2:1$. This is illustrated in Fig.\ 2b.
Regarding $s_\parallel$ as the softening length of reference, different
values of $s_\perp/s_\parallel$ have no influence on the planar functions and
change the gravitational interaction along the vertical direction according
to simple scaling laws.

\section{Discussion}

Three recent papers concerning softening optimization and conception deserve
comment:
\begin{itemize}
\item Merritt's (1996) optimization is performed with respect to the
      Newtonian dynamics in the configuration space, and concerns the
      softening length. The configuration space does not permit a clear
      distinction between large-scale dynamical properties and small-scale
      noise, and also emphasizes the equilibrium state as most representative
      of the whole dynamics.
\item Weinberg's (1996) optimization is comparable to that of Merritt (1996)
      but concerns orthogonal series force computation, i.e.\ roughly
      speaking the softening length and the type of softened gravity.
\item Dyer \& Ip's (1993) conception is rigid: softened gravity must mimic
      finite-sized particles. But why? Softening is an artifice: its physical
      consistency should be scrutinized with respect to basic dynamical
      requirements, not with respect to the inter-particle force alone. An
      elastic conception is more useful. Apart from that, Dyer \& Ip (1993)
      have suggested a softening optimization that is not so different from
      that of Merritt (1996).
\end{itemize}

\section{Conclusions and perspectives}

The importance of computer simulations in astrophysics is analogous to that
of experiments in other branches of physics. They also serve as a welcome
bridge between theories, often restricted to idealized situations, and
observations, revealing instead the complexity of nature. Major present
objectives are to construct physically consistent $N$-body models of disc
galaxies and to simulate their dynamical evolution, especially in regimes of
spiral structure in which a fruitful comparison between theories and
simulations can be made (e.g., Pfenniger \& Friedli 1991; Junqueira \& Combes
1996; Zhang 1996; Bottema \& Gerritsen 1997; Fuchs \& von Linden 1998; von
Linden et al.\ 1998; Zhang 1998a, b). The construction of such models is
indeed a difficult task which has not yet been fully accomplished, and which
should eventually provide clues of vital importance to a number of open
questions posed by both theories and observations.

   Our involvement has been threefold. In Paper I, we have recognized a
fundamental problem posed by this research programme (for a concrete use of
that analysis and for interesting remarks see, e.g., Junqueira \& Combes
1996). In Paper II, we have devised a method for solving this problem. In the
present paper, we apply this method and solve the problem, thus laying the
foundations of such a plan. The {\it major result\/} is that gravity can be
modelled so as to optimize the fidelity of simulations, and the procedure is
practicable. The following conclusions point up the whys and wherefores:
\begin{enumerate}
\item Optimization is performed with respect to arbitrary dynamical
      requirements and, in specific examples, with respect to the Newtonian
      dynamics. This enriches the modelling with an {\it unprecedented\/}
      degree of freedom, which has clear epistemological motivations (cf.\
      Sect.\ 1, discussion of the present paper).
\item Optimization is performed in the wavenumber space. This is the {\it
      appropriate\/} domain for imposing dynamical requirements on the
      modelling.
\item Optimization concerns {\it both\/} the softening length {\it and\/} the
      type of softened gravity.
\item Softening is conceived as a {\it double\/} artifice. The softened
      gravity and finite-sized particle conceptions are equivalent in the
      basic cases. Concerning more complex situations, the latter is
      particularly useful for implementing star-gas softening, whereas the
      former is particularly useful for implementing anisotropic softening.
      Thus both conceptions contribute towards the accomplishment of a
      physically consistent modelling.
\end{enumerate}
Our application is ready for a concrete use. An attractive idea is to employ
a particle-particle code together with MD-GRAPE, a highly parallelized
special-purpose computer for many-body simulations with an arbitrary central
force (Fukushige et al.\ 1996). We can also employ a classical particle-mesh
code. Then the dynamical effects of the grid are known and factorize as those
of softening (e.g., Bouchet et al.\ 1985; Efstathiou et al.\ 1985; for a
review see, e.g., Hockney \& Eastwood 1988). So essentially the application
proceeds as in the present paper, but it may be useful to act directly on the
wavenumber space (e.g., Tormen \& Bertschinger 1996). A more complex problem
concerns tree codes, which have hierarchical structure and adaptive
resolution over multiple scales (e.g., Hernquist 1987; for a review see,
e.g., Pfalzner \& Gibbon 1996). The solution to that problem would need a
more advanced analysis (cf.\ following discussion). Welcome suggestions about
the choice of the code can come from cosmological simulations (e.g., Splinter
et al.\ 1998).

   Finally, what about the future? Our approach is connected with the
technique of filtering in spectral domain used in the context of digital
image processing. This is a rapidly evolving field with growing applications
in science and engineering, which can promote further substantial advances in
$N$-body modelling of disc galaxies. For instance, wavelets are ideal for
resolving multi-scale problems in space and/or time, such as those concerning
turbulence, bifurcations, fractals and many others (see, e.g., Kaiser 1994;
Holschneider 1995; Bowman \& Newell 1998; for an alternative analysis tool
see, e.g., Stutzki et al.\ 1998). Speculating further, wavelets might be used
for speeding up simulations through fast solution of linear systems (cf.\
Press et al.\ 1992, pp.\ 597--599 and 782).

   These are the merits of our contribution. We hope that the trilogy (Papers
I--III) and further reflections (Romeo 1998) will encourage $N$-body
experimenters to model gravity so as to optimize the fidelity of their
simulations, and that the result will be a stronger interdisciplinary
connection with theories and observations.

\begin{acknowledgements}

This paper is dedicated to my children Johan, Filip and Lukas, in the hope
that they can always open their mind towards new horizons. It is a great
pleasure to thank John Black, Francoise Combes, Lars Hernquist, Cathy
Horellou and Daniel Pfenniger for strong encouragement and valuable
suggestions on a previous draft of this paper, and an anonymous referee for
positive comments. In addition, I am very grateful to Neil Comins, Horacio
Dottori, Daniel Friedli, Dick Miller and Michel Tagger for strong
encouragement and useful discussions. Besides, I am very thankful to Marek
Abramowicz for his invaluable help.

\end{acknowledgements}

\appendix

\section{ABC of finite-sized particles}

Finite-sized particles interacting with Newtonian gravity are analogous to
point particles interacting with softened gravity. The dynamics of
one-component 2-D discs containing such particles can be investigated by
performing an analysis comparable to that of Sects.\ II-2.1 and 2.1. In this
appendix we report the formulae useful for Sect.\ 2.2. Let $m\mu_s(R)$ be the
particle mass distribution of scale length $s$. The reduction factor is:
\begin{equation}
{\cal S}(|k|s)=\{2\pi\,\,{\cal H}_0[\mu_s(R)](k)\}^2\, .
\end{equation}
The inversion formula for $\mu_s(R)$ is:
\begin{equation}
\mu_s(R)
=\frac{1}{2\pi}\,\,{\cal H}_0\left[\sqrt{{\cal S}(|k|s)}\,\right](R)\, .
\end{equation}
Last and most useful, the inversion formulae for the interaction potential
$-Gm^2\varphi_s(R)$ and force $-Gm^2f_s(R)$ are:
\begin{equation}
\varphi_s(R)={\cal H}_0\left[\frac{1}{k}\,{\cal S}(|k|s)\right](R)\, ,
\end{equation}
\begin{equation}
f_s(R)={\cal H}_1[{\cal S}(|k|s)](R)\, .
\end{equation}
(For general references about finite-sized particles see, e.g., Vlasov 1961;
Rohrlich 1965; Dawson 1983; Hockney \& Eastwood 1988; Birdsall \& Langdon
1991; in particular, the last reference contains useful insights in the
context of $N$-body simulations of plasmas.)

\end{document}